\documentstyle[aps,prl,multicol,amsmath,epsf,graphics]{revtex}

\begin{document}

\draft

\title{Exact solutions for diluted spin glasses and optimization problems}

\author{
Silvio Franz$^1$, Michele Leone$^{1,2}$, Federico Ricci-Tersenghi$^1$
and Riccardo Zecchina$^1$
}

\address{
$^1$ ICTP, Condensed Matter Group, P.O. Box 586, I-34014 Trieste,
Italy\\
$^2$ SISSA, Via Beirut 9, I-34014 Trieste, Italy
}

\date{\today}

\maketitle

\begin{abstract}
We study the low temperature properties of p-spin glass models with
finite connectivity and of some optimization problems.  Using a
one-step functional replica symmetry breaking Ansatz we can solve
exactly the saddle-point equations for graphs with uniform
connectivity. The resulting ground state energy is in perfect
agreement with numerical simulations.  For fluctuating connectivity
graphs, the same Ansatz can be used in a variational way: For $p$-spin
models (known as $p$-XOR-SAT in computer science) it provides the
exact configurational entropy together with the dynamical and static
critical connectivities (for $p=3$, $\gamma_d=0.818$ and
$\gamma_s=0.918$ resp.), whereas for hard optimization problems like
3-SAT or Bicoloring it provides new upper bounds for their critical
thresholds ($\gamma_c^{var}=4.396$ and $\gamma_c^{var}=2.149$ resp.).
\end{abstract}


\vspace{-.2cm}

\begin{multicols}{2}
\narrowtext

The nature of the glassy phase and of the out-of-equilibrium dynamics
of physical systems are two intertwined aspects of the behavior of
many complex systems found in different fields, ranging from physics
or biology to computer science and game theory.  The existence and the
characterization of long time states, and the question of relaxation
times are important open issues of modern statistical mechanics and
probability theory.

Fully connected spin glasses have served as prototype models able to
provide a highly non trivial static and off-equilibrium phenomenology
already at the mean-field level~\cite{MEPAVI,YOUNG_BOOK}.  However,
some important features of complex real-world systems heavily rely on
the connectivity pattern which is particularly simple in these
systems.

For instance, super-cooled liquids and structural glasses are
characterized by a finite number of short range interactions for each
particle (bounded by the so called {\it kissing number} of the
particles), which leads to a complex structure of energy and entropy
barriers in phase space. Heterogeneities in both the static and the
off-equilibrium regimes witness such underlying
constraints~\cite{ANGELL}.

In computer science, non-trivial ensembles of hard combinatorial
optimization problems, the so called NP-complete
problems~\cite{GAREY}, typically map onto spin glasses with finite
average connectivity (or degree) at zero temperature~\cite{MAP}.  In
the last years methods from statistical physics have been very useful
in order to study phase transitions in such problems~\cite{TRANS}.
The hardest among these share characteristics that are largely
independent on the specific algorithms adopted for their
solution. Important features such as solution time or memory
requirements are conjectured to be strictly related to the geometrical
structure of their low temperature phase space~\cite{NATURE}.  For a
survey on the state of the art we address the reader to two recent
special issues~\cite{SPECIAL}.

The main technical obstacle for the development of a statistical
physics theory over finite degree graphs has been the extension of the
Parisi replica symmetry breaking (RSB) scheme to the functional
level~\cite{DILUTED,WOSH}. The replica symmetric (RS) phases are
described by a single probability distribution function of the
effective fields, capturing the site-to-site fluctuations of the local
magnetization. On the contrary, when the symmetry among replicas
breaks down, there appear many pure states, each one endowed with its
own effective field distribution.  The site-to-site fluctuations
induce correlations among such probability distributions. The overall
free energy has to be optimized in a large functional space which
becomes more and more complex as the Parisi scheme is
iterated. Recently in~\cite{MEPA} a population dynamics algorithm has
been introduced which is able, for large enough computer resources, to
reconstruct the field distributions that lead to the numerical
solution of the one-step RSB (1RSB) equations in full generality.

In this Letter we re-examine in the appropriate 1RSB context an Ansatz
previously introduced in~\cite{WOSH} to get an approximated solution
of the Viana-Bray model.  That Ansatz, which neglects site-to-site
fluctuations, allows for great simplifications in the functional
equations describing the low temperature regime of spin glasses and
optimization problems defined on graphs with finite connectivity. The
Ansatz solves exactly the saddle-point equations for models defined
over uniform degree random hypergraphs and it is simple enough to
allow for the explicit computation of the thermodynamic quantities at
low temperature, which remarkably agree with numerics.  As
representative instances, we compute the ground state (GS) energy of
the $p$-spin glass and of the Bicoloring problem. In both cases, the
problem of finding the GS is a hard computational task (NP-hard).  To
the best of our knowledge, the one presented here is the first exact,
fully analytical solution for diluted models in the RSB phase.

For the $p$-spin glass with fluctuating connectivities where
site-to-site correlations are in principle important, the same Ansatz
allows to evaluate exactly the configurational entropy and the
dynamical and static transition points found as the average
connectivity is increased~\cite{RIWEZE}.  In computer science this
model is known as random $p$-XOR-SAT~\cite{XOR-SAT} and its complete
probabilistic characterization is considered an open problem (at
present only lower and upper rigorous bounds to the critical threshold
are known~\cite{CREDADU}).  From a combinatorial standpoint this
problem has straight connections with the well-studied domain of
random linear systems over finite fields, with applications in coding
and cryptography~\cite{CODING}.

In the fluctuating connectivity framework, we have also tested the
method as a variational approach to the study of the GS properties of
random NP-complete combinatorial problems, such as random 3-SAT or
Bicoloring~\cite{GAREY}, obtaining the currently most accurate
analytical estimation of their critical SAT/UNSAT thresholds (see below).

In what follows we shall concentrate on the $p=3$ spin interactions,
the generalization to arbitrary $p$ being straightforward.  The
Hamiltonian of the models we have chosen to study reads $H =
\sum_{[i,j,k] \in E} G[S_i,S_j,S_k]$, where $S_i=\pm1$ are Ising spins
and $E$ is the set of triples (plaquettes), which form a random
hypergraph (locally the graph has the topology of a Husimi tree).  In
the fixed connectivity case, every index $i=1,\ldots,N$ must appear in
$E$ the same number $k\!+\!1$ of times.  However the presence of
hyperloops~\cite{RIWEZE} of length of the order $\log(N)$ induce non
trivial contributions to the free-energy.  In the fluctuating
connectivity case, each possible plaquette is chosen at random with
probability $\gamma/N^2$.

In the 3-spin glass case, the local interaction reads $G[S_i,S_j,S_k]= -
J_{ijk} S_i S_j S_k$ with $J_{ijk}=\pm1$ randomly.  For the Bicoloring
problem we have $G[S_i,S_j,S_k ]= \delta(S_i,S_j) \delta(S_i,S_k)=
\frac14 (S_i S_j + S_i S_k+ S_i S_j+1)$. Each interaction adds zero
energy only if the spin configuration is not monochromatic, i.e. not all
spins on the plaquette are equal. The optimization problem amounts at
minimizing the number of monochromatic plaquettes.

Firstly we focus on the fixed connectivity case (with connectivity
$k\!+\!1$).  In order to calculate the averaged free energy at inverse
temperature $\beta$ we resort to the replica method, where the average
free energy is evaluated from an analytic continuation of the integer
moments of the partition function~\cite{MEPAVI}. The replicated free
energy reads
\begin{multline}
\beta f_n = \overline{Z^n} =
\underset{f(\vec\sigma)}{\mbox{extr}} \Bigg[ k \ln\left( 
\sum_{\vec\sigma} f(\vec\sigma)^{\frac{k+1}{k}} \right) -
{k\!+\!1 \over 3} \\
\ln \Bigg(
\sum_{\vec\sigma_1,\vec\sigma_2,\vec\sigma_3} f(\vec\sigma_1)
f(\vec\sigma_2) f(\vec\sigma_3) e^{\beta \sum_{a=1}^n {\cal
G}[\sigma_1^a,\sigma_2^a,\sigma_3^a]} \Bigg) \Bigg] ,
\label{eq:fn}
\end{multline}
with ${\cal G}[x,y,z]= x y z$ for the spin glass and ${\cal G}[x,y,z]=
-(x y +y z +x z+1)/4$ for Bicoloring.  $a=1,..,n$ is the replica
index, $\vec\sigma=\{\sigma^1,..,\sigma^n\}$ is a vector of $n$ Ising
variables and $f(\vec\sigma) \propto c(\vec\sigma)^k$, where
$c(\vec\sigma) = N^{-1} \sum_i \delta(\vec S_i-\vec\sigma)$ counts the
fraction of sites $i$ having replicated spin $S_i^a=\sigma^a$.  The
functional order parameter $f(\vec\sigma)$ must be symmetric in
$\vec\sigma$.  From Eq.(\ref{eq:fn}) one obtains the following saddle
point equations
\begin{eqnarray}
f(\vec\sigma) & = & \frac{D^k(\vec\sigma)} {\sum_{\vec\sigma}
D^k(\vec\sigma)} \quad ,
\label{eq:spe1}\\
D(\vec\sigma) & = & \sum_{\vec\tau,\vec\mu} f(\vec\tau) f(\vec\mu)
\exp \left( \beta \sum_{a=1}^n {\cal G}[\sigma^a, \tau^a, \mu^a]
\right) \quad .
\label{eq:spe2}
\end{eqnarray}
The above equations admit a paramagnetic and a spin glass solution.
As long as replica symmetry holds (e.g.\ in the paramagnetic or
ferromagnetic phases) the order parameter $f(\vec\sigma)$ depends only
on the sum $\sum_a \sigma^a$, i.e.\ on a single probability
distribution of effective local fields.  However, in the glassy phase
the RS solution is not optimal (and yet stable), and one needs a RSB
Ansatz, which is in general very complicated, even in the simplest
case of 1RSB.  Nevertheless in our case, the observation that the
sites are locally equivalent suggests to neglect site-to-site
fluctuations in the distribution of the effective fields.  In the
replica formalism this fact is reflected in the use of the following
{\em factorized} Ansatz~\cite{WOSH,BETA}
\begin{equation}
f(\vec\sigma) = \prod_{g=1}^{n/m} \tilde{f}(\vec\sigma_g) =
\prod_{g=1}^{n/m} \int dh\: P(h)\, \frac{e^{\beta h \sum_{a=1}^m
\sigma_g^a}}{(2 \cosh\beta h)^m} \ ,
\label{eq:factor}
\end{equation}
where the $n$ replicas have been divided into $n/m$ groups of $m$
replicas each and $\vec\sigma_g$ is an $m$-dimensional vector with the
components belonging to the $g$-th group.  The general interpretation
of the 1RSB order parameter~\cite{MEPA} shows that $P(h)$ is the
probability distribution of the cavity field, i.e.\ the field on a
site after one of the interactions of that site has been removed.  The
above Ansatz is consistent with the saddle point equations
(\ref{eq:spe1},\ref{eq:spe2}), and the very same equations are
verified by $\tilde{f}(\vec\sigma_g)$ and $\tilde{D}(\vec\sigma_g)$
with the sum in (\ref{eq:spe2}) running only up to $m$.

The RSB saddle point equations are different for the two models.
In the 3-spin case we find
\begin{equation}
\frac{P(h)}{(2 \cosh \beta h)^m} = A_k^{-1} \int \prod_{i=1}^k
\frac{du_i\, Q(u_i)}{(\cosh \beta u_i)^m}\,\delta\!\left(h -
\sum_{i=1}^k u_i\right),
\label{eq:ph1}
\end{equation}
where $Q(u) = \int {\cal D}h {\cal D}g \, \delta[u - u(h,g)]$, with
${\cal D}h = dh\,P(h)$ and $\tanh[\beta u(h,g)] = \tanh(\beta)
\tanh(\beta h) \tanh(\beta g)$, and $A_k$ normalizes the $P(h)$.  For
Bicoloring we find
\begin{eqnarray}
\frac{P(h)}{(2 \cosh \beta h)^m} & = & A_k^{-1} \int \prod_{i=1}^k
\frac{{\cal D}h_i\,{\cal D}g_i\,W(h_i,g_i)^m} {(4 \cosh(\beta h_i)
\cosh(\beta g_i))^m} \cdot \nonumber \\
 & & \cdot \; \delta\!\left(h - \sum_{i=1}^k t(h_i,g_i)\right) \quad ,
\label{eq:ph2}
\end{eqnarray}
where $t(h,g)=\frac12 \beta^{-1} \ln(\alpha_{-}/\alpha_{+})$ and
$W(h,g)=\sqrt{\alpha_- \alpha_+}$ with $\alpha_{\pm} = 2
\cosh[\beta(h-g)] + 2 \cosh[\beta (h+g \pm 1)] e^{-\beta}$.  The RS
(resp. paramagnetic) equation is recovered for $m=n$ (resp. $m=1$).
As usual, in order to find the thermodynamical free-energy a
maximization of the free-energy functional with respect to $m$ should
be performed.

For a generic temperature the solutions to (\ref{eq:ph1}) and
(\ref{eq:ph2}) can be easily found numerically with a RS-like
population dynamics, which requires much less computational effort
than the 1RSB algorithm of Ref.~\cite{MEPA}.

Interestingly enough in the limit of zero temperature we can solve the
equations analytically.  Indeed for $\beta \to \infty$ we have $u(h,g)
\to {\rm sign}(h\,g) \min(1,h,g)$ and $t(h,g) \to -{\rm sign}(h)
\min(1,h,g)\,\theta(h\,g)$ (where $\theta$ is the step function).
Then both $u(h,g)$ and $t(h,g)$ can take only the values $0$ and $\pm
1$ for integer valued cavity fields.  Rational valued solutions also
exist and yet vanish close to the free energy maximum.

For the $p$-spin with odd connectivity $k+1$ and for the Bicoloring
the analytical expressions are quite involved and will be given
elsewhere~\cite{FLRZ}.  For $p$-spin with even connectivities
(i.e.\ odd $k$) the solution can be written in a very compact way,
\begin{gather}
f_0(y,k\!+\!1) = \frac{2k-1}{3} g(y,k\!+\!1)
- \frac{2k+2}{3} g(y,k) \quad , \nonumber \\
\textnormal{with} \quad g(y,k) = \frac1y \ln\left[ 2^{-k} \sum_{i=0}^k
\binom{k}{i} e^{y\,|k-2i|} \right] \quad ,
\end{gather}
where $y$ is the zero temperature limit of the quantity $\beta m$
which turns out to be finite.  The GS energy ($e_{gs}$) corresponds to
the maximum of $f_0$~\cite{MEPAVI}.  For connectivities smaller than 4
for the $p$-spin and 7 for the Bicoloring the maximum is always in
$y=\infty$ and corresponds to the RS paramagnetic solution
$P(h)=\delta(h)$.  The RS spin glass solution, located in $y=0$, has
always a lower energy with respect to the physical one.  For some
connectivities the free energy values ($e_{gs}$) are reported in the
tables, together with the corresponding $y=y^*$ saddle-point values.
For the 3-spin case we also report numerical estimations of the GS
energy ($e_{gs}^{num}$) obtained by extrapolating the results of
exhaustive enumerations (sizes up to $N=60$ averaged over 1000--10000
samples).  Moreover in~\cite{FMRWZ} the $y^*$ value for the 3-spin
model with $k+1=4$ has been estimated to be 1.41(1), perfectly
compatible with our analytic value.

A further check of the analytic solution in the $p$-spin case is
provided by the proper convergence, in the limit $k \to \infty$, to
the exact solution of the fully connected $p$-spin model by
Gardner~\cite{GARDNER} after a proper rescaling of the coupling.

\begin{center}
\begin{tabular}{||c|c|}\hline
\ \rotatebox{90}{3 - spin} \ &
\begin{tabular}{|c|c|c|c|}\hline
\ $k$+1\ \ & $e_{gs}$ & $e_{gs}^{num}$ & $y^*$ \\ \hline \hline
  1--3     &\ -($k$+1)/3\ \ &               & $\infty$    \\ \hline
   4       &   -1.21771     &\ -1.218(6)\ \ &\ 1.41155\ \ \\ \hline
   5       &   -1.39492     &  -1.395(7)    &  1.09572    \\ \hline
   6       &   -1.54414     &  -1.544(9)    &  0.90163    \\ \hline
\end{tabular}
\\ \hline
\end{tabular}
\end{center}

\begin{center}
\begin{tabular}{||c|c|}\hline
\ \rotatebox{90}{Bicoloring} \ &
\begin{tabular}{|c|c|c|}\hline
\ $k$+1\ \ & $e_{gs}$ &  $y^*$      \\ \hline \hline
 1--6  &     0        & $\infty$    \\ \hline
  7    &\ 0.003711\ \ &\ 1.96611\ \ \\ \hline
  8    &  0.027383    & 1.17118     \\ \hline
  9    &  0.058131    & 0.92887     \\ \hline
  10   &  0.093181    & 0.78746     \\ \hline
  11   &  0.131392    & 0.69315     \\ \hline
  12   &  0.172118    & 0.55338     \\ \hline
\end{tabular}
\\ \hline
\end{tabular}
\end{center}

In the case of models defined over non-homogeneous graphs the Ansatz
can be used to obtain approximate or, in some cases, exact variational
estimates of the thermodynamic functions.  We have considered three
representative cases: the 3-spin model and the Bicoloring problem over
random hypergraphs with a Poisson distribution of site connectivities
and the random 3-SAT problem.

\begin{figure}
\epsfxsize=0.95\columnwidth
\epsffile{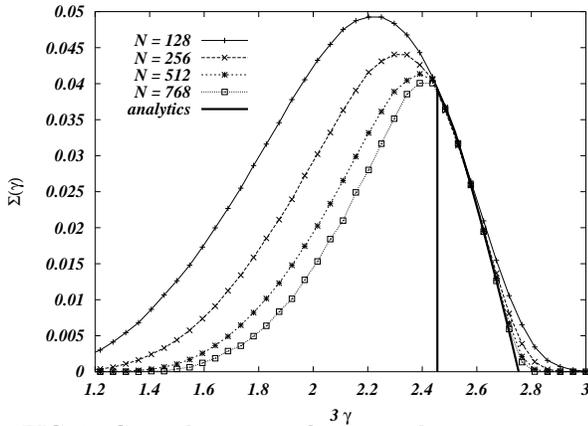}
\caption{Ground states configurational entropy versus mean
connectivity for the 3-spin.  The results of numerical clustering with
an overlap cut-off of 0.7 (averaged over 1000, 1000, 500 and 50
samples) converge to the analytical prediction.}
\label{fig}
\end{figure}

For the 3-spin model we obtained estimates of the dynamical and static
critical points looking at the {\em configurational entropy}.  It has
been shown in~\cite{RIWEZE} that in this model, the GS are
clustered. Given a GS there is an exponential number of other GS which
can be reached through GS paths where subsequent GS differ only by a
finite number of spin flips. For small average connectivity there is a
unique cluster, while above a threshold $\gamma_d$ the number of
disconnected cluster become exponentially large. The configurational
entropy is the logarithm of the number of cluster per spin, and can be
computed in the replica 1RSB formalism as $\Sigma(\gamma) =
\left.\frac{\partial f}{\partial m}\right|_{m=1}$~\cite{REMI}.
$\Sigma(\gamma)$ jumps to a non zero value at the dynamical critical
point $\gamma_d$ and then it vanishes again at the static critical
point $\gamma_s$.  Using arguments put forward in~\cite{FP}, one can
show~\cite{FLRZ} that, due to the triviality of the paramagnetic phase
of the model~\cite{RIWEZE}, where $P(h)=\delta(h)$, the factorized
Ansatz (\ref{eq:factor}) yields the exact result.  The resulting
expression reads
\begin{equation}
\Sigma(\gamma) = \ln(2) [r - 3 \gamma r^2 (1-r) - \gamma r^3] \quad ,
\label{eq:complexity}
\end{equation}
with $r$ solving the $m=1$ saddle point equation $1-r=\exp(-3\gamma
r^2)$.  The above expression is different from zero between
$\gamma_d\!=\!0.818$ and $\gamma_s\!=\!0.918$ (see bold line in
Fig.~\ref{fig}) where it equals the difference between the
paramagnetic and the ferromagnetic entropies~\cite{RIWEZE}.  The above
critical points coincide with numerical estimates~\cite{RIWEZE}.

In this model the static RSB transition point $\gamma_s$ coincides
with the critical point $\gamma_c$ beyond which the system becomes
frustrated, that is, no longer all interactions can be satisfied at
the same time and therefore the GS energy becomes
positive~\cite{RIWEZE}.  In computer science this point is known as
the SAT/UNSAT critical threshold: $\gamma_c=\gamma_s=0.918$ thus
provides the critical threshold for random 3-XOR-SAT, in perfect
agreement with the numerical estimation~\cite{RIWEZE}.

Above $\gamma_s$ the factorized Ansatz ceases to be exact and can be
used only at a variational level.

In order to check numerically Eq.(\ref{eq:complexity}), we have
performed a GS clustering.  This task is in general very hard due to
the large number of GS and because of the lack of a proper definition
of clusters in a finite size systems.  However in this model
calculations are easier, thanks to the presence of only two relevant
overlaps between GS: The internal overlap equals $r$ and is always
larger than 0.7, while different clusters are almost orthogonal.  This
leads to an optimal cut-off of 0.7 for the numerical identification of
clusters.  The results shown in Fig.~\ref{fig} are in remarkable
agreement with the analytical curve.

Finally, in models where the configurational entropy is likely to be
zero and correlation among clusters are stronger, like Bicoloring over
fluctuating degree random graphs and 3-SAT~\cite{BIMOWE}, the dynamic
and static critical points coincide ($\gamma_d=\gamma_s$) and precede
the SAT/UNSAT critical point $\gamma_c$ (usually called $\alpha_c$).
Though less effective, the factorized Ansatz with integer valued
fields still provides an estimate for $\gamma_c$~\cite{nota_bound}.
Namely $\gamma^{var}_c = 4.3962$ for 3-SAT and $\gamma^{var}_c =
2.149$ for Bicoloring, which improve present rigorous bounds (recently
reviewed in~\cite{BOUNDS}).

Summarizing, we have studied a factorization Ansatz which allows to
solve exactly diluted spin-glass and optimization models, on
homogeneous hypergraphs.  These analytical results can play a crucial
role in testing convergence of heuristic algorithms.  For
non-homogeneous graphs the Ansatz still allows to get very good
results (some of them exact) as we have shown for the $p$-spin, the
Bicoloring and the 3-SAT problems.  Encouraged by the recent rigorous
results obtained on simpler models in~\cite{2-SAT,TALAGRAND}, we trust
that an alternative and mathematically rigorous derivation of our
results may be possible.

We thank M. M\'ezard, G. Parisi and M. Weigt for several useful
discussions.

\end{multicols}


\vspace{-.5cm}

\begin{references}

\vspace{-1.5cm}

\bibitem{MEPAVI} M. M\'ezard, G. Parisi and M.A. Virasoro, {\it Spin
Glass Theory and Beyond} (World Scientific, Singapore, 1987).

\bibitem{YOUNG_BOOK} {\it Spin Glasses and Random Fields}, A.P.~Young
Ed. (World Scientific, Singapore, 1998).

\bibitem{ANGELL} C.A. Angell, Science {\bf 267}, 1924 (1995).

\bibitem{GAREY} M. Garey and D.S. Johnson, {\it Computers and
Intractability; A guide to the theory of NP-completeness} (Freeman,
San Francisco, 1979); C.H.~Papadimitriou, {\it Computational
Complexity} (Addison-Wesley, 1994).

\bibitem{MAP} Y.T. Fu and P.W. Anderson, in {\it Lectures in the
Sciences of Complexity}, D. Stein Ed. (Addison-Wesley, 1989), p.~815.
R. Monasson and R. Zecchina, Phys. Rev. Lett. {\bf 76}, 3881 (1996);
Phys. Rev. E {\bf 56}, 1357 (1997).

\bibitem{TRANS} S. Mertens, Phys. Rev. Lett. {\bf 81}, 4281
(1998). M. Weigt and A.K. Hartmann, Phys. Rev. Lett. {\bf 84}, 6118
(2000).

\bibitem{NATURE} R. Monasson, R. Zecchina, S. Kirkpatrick, B. Selman
and L. Troyansky, Nature (London) {\bf 400}, 133 (1999).

\bibitem{SPECIAL} {\it ``Statistical Physics Methods in Discrete
Probability, Combinatorics, and ...''}, J.T. Chayes and D. Randall
Edts., Random Struct. Algor. {\bf 15} (1999). {\it ``NP-hardness and
Phase Transitions''}, O. Dubois, R. Monasson, B. Selman, R. Zecchina
Edts., Theor. Comp. Science in press.

\bibitem{DILUTED} L. Viana and A.J. Bray, J. Phys. C {\bf 18}, 3037
(1985); D.J.~Thouless, Phys. Rev. Lett. {\bf 56}, 1082 (1986);
I.~Kanter and H.~Sompolinsky, Phys. Rev. Lett. {\bf 58}, 164 (1987);
M.~M\'ezard and G.~Parisi, Europhys. Lett. {\bf 3}, 1067 (1987);
C.~De~Dominicis and P.~Mottishaw, J. Phys. A {\bf 20}, L1267 (1987);
J.M.~Carlson, J.T.~Chayes, L.~Chayes, J.P.~Sethna and D.J.~Thouless,
Europhys. Lett. {\bf 5}, 355 (1988); Y.Y.~Goldschmidt and P.Y.~Lai,
J. Phys. A {\bf 23}, L775 (1990); H. Rieger and T.R. Kirkpatrick,
Phys. Rev. B {\bf 45}, 9772 (1992); R.Monasson, J.Phys.A {\bf 31},
513(1998).

\bibitem{WOSH} K.Y.M.Wong, D.S.Sherrington, J.Phys.A {\bf 21},L459(1988).

\bibitem{MEPA} M. M\'ezard and G. Parisi, Eur. Phys. J. B {\bf 20},
217 (2001).

\bibitem{RIWEZE} F. Ricci-Tersenghi, M. Weigt, R. Zecchina,
Phys. Rev. E {\bf 63}, 026702 (2001).

\bibitem{XOR-SAT} T.J. Schaefer, in Proc. 10th STOC, San Diego (CA,
USA), ACM, 1978, p.216.

\bibitem{CREDADU} N. Creignou, H. Daud\'e and O. Dubois, preprint {\tt
arXiv:cs.DM/0106001}.

\bibitem{CODING} R.G. Gallager, {\it Low Density Parity Check Codes}
(MIT Press, 1963). I. Kanter and D. Saad, Phys. Rev. Lett. {\bf 83},
2660 (1999). Y. Kabashima, T. Murayama and D. Saad,
Phys. Rev. Lett. {\bf 84}, 1355 (2000); Phys. Rev. Lett. {\bf 84},
2030 (2000).

\bibitem{FMRWZ} S. Franz, M. M\'ezard, F. Ricci-Tersenghi, M. Weigt
and R. Zecchina, preprint {\tt cond-mat/0103026}.

\bibitem{BETA} Through a rescaling of $\beta \to 2 \beta$ for the
Bicoloring.

\bibitem{FLRZ} S. Franz, M. Leone, F. Ricci-Tersenghi and R. Zecchina,
in preparation.

\bibitem{GARDNER} E. Gardner, Nucl. Phys. B {\bf 257}, 747 (1985).

\bibitem{REMI} R. Monasson, Phys. Rev. Lett. {\bf 75}, 2817 (1995).

\bibitem{FP} S. Franz and G. Parisi, J. Phys. I (France) {\bf 5}, 1401
(1995).

\bibitem{BIMOWE} G. Biroli, R. Monasson, M. Weigt, Eur. Phys. J. B
{\bf 14}, 551 (2000).

\bibitem{nota_bound} In the replica method one maximizes the free
energy functional, thus providing a upper bound to the true
$\gamma_c$.

\bibitem{BOUNDS} D. Achlioptas, Theor. Comp. Science to appear.

\bibitem{2-SAT} B. Bollob\`as, C. Borgs, J. Chayes, J.H. Kim,
D.B. Wilson, Random Struct. Algor. {\bf 18}, 201 (2001).

\bibitem{TALAGRAND} M. Talagrand, Prob. Theory Rel. Fields {\bf 117},
303 (2000).

\end{references}
\end{document}